\documentclass[preprint,3p,times,sort&compress]{elsarticle}

\usepackage{multirow,setspace,times,amssymb,amsmath,graphicx,color,rotating,subfigure,url}
\usepackage{lineno,color}
\usepackage{natbib}
\usepackage{booktabs}%
\graphicspath{{Figures/}}
\usepackage[table]{xcolor}
\usepackage{tabularx}
\usepackage{graphicx} 
\usepackage{epstopdf}
\usepackage{mathrsfs}
\usepackage{makecell}
\usepackage[bookmarks=true,colorlinks,linkcolor=blue,anchorcolor=blue,citecolor=blue,unicode]{hyperref}
\usepackage{bookmark}
\usepackage{ulem}

\hypersetup{CJKbookmarks=true}%
\journal{Energy}
\begin{document}

\begin{frontmatter}
\title{Resilience of international oil trade networks under extreme event shock-recovery simulations}

\author[SC]{Na Wei}
\author[SB,RCE]{Wen-Jie Xie}
\author[SB,RCE,DM]{Wei-Xing Zhou\corref{coa}}
\ead{wxzhou@ecust.edu.cn}
\cortext[coa]{Corresponding author. Corresponding to: 130 Meilong Road, P.O. Box 114, School of Business, East China University of Science and Technology, Shanghai 200237, China.}

\address[SC]{School of Civil and Architectural Engineering, Yangtze Normal University, Chongqin 408100, China}
\address[SB]{School of Business, East China University of Science and Technology, Shanghai 200237, China}
\address[RCE]{Research Center for Econophysics, East China University of Science and Technology, Shanghai 200237, China}
\address[DM]{School of Mathematics, East China University of Science and Technology, Shanghai 200237, China}

\begin{abstract}

With the frequent occurrence of black swan events, global energy security situation has become increasingly complex and severe. Assessing the resilience of the international oil trade network (iOTN) is crucial for evaluating its ability to withstand extreme shocks and recover thereafter, ensuring energy security. We overcomes the limitations of discrete historical data by developing a simulation model for extreme event shock-recovery in the iOTNs. We introduce network efficiency indicator to measure oil resource allocation efficiency and evaluate network performance. Then, construct a resilience index to explore the resilience of the iOTNs from dimensions of resistance and recoverability. Our findings indicate that extreme events can lead to sharp declines in performance of the iOTNs, especially when economies with significant trading positions and relations suffer shocks. The upward trend in recoverability and resilience reflects the self-organizing nature of the iOTNs, demonstrating its capacity for optimizing its own structure and functionality. Unlike traditional energy security research based solely on discrete historical data or resistance indicators, our model evaluates resilience from multiple dimensions, offering insights for global energy governance systems while providing diverse perspectives for various economies to mitigate risks and uphold energy security.

\end{abstract}

\begin{keyword}
 International oil trade network; Network resilience; Oil security; Trade shock simulation
\\
\end{keyword}

\end{frontmatter}


\section{Introduction}


Former US Secretary of State Kissinger once stated, "whoever controls oil controls all countries." As the backbones of an economies, oil significantly impacts global politics, economic order, and military activities, playing a crucial role in international economic security. The unbalance distribution of oil resources globally has led to a disconnect between supply and consumption markets, making global oil trade essential for balancing supply and demand  \cite{Xi-Zhou-Gao-Liu-Zheng-Sun-2019-EnergyEcon,Caraiani-2019-EnergyEcon,Sun-An-Gao-Guo-Wang-Liu-Wen-2019-Energy}. However, in recent years, frequent occurrences of 
black swan incidents have disrupted the global energy environment as well as social and ecological landscapes. Events such as the global financial crisis triggered by the subprime mortgage crisis and the worldwide spread of novel coronavirus pneumonia have been followed by events like the Russia-Ukraine conflict and international environmental turbulence due to extreme weather conditions. These incidents have severely impacted international commodity trade including crude oil. Consequently, the international oil market is becoming increasingly uncertain. The inherent energy market risks associated with navigating uncertainty within the oil trade market have emerged as 
the central concern for various economies to uphold both national and energy security \cite{An-Wang-Qu-Zhang-2018-Energy,Du-Wang-Dong-Tian-Liu-Wang-Fang-2017-AEn,Du-Dong-Wang-Zhao-Zhang-Vilela-Stanley-2019-Energy}.


The interconnectedness of global economies means that changes in market conditions within one region can swiftly impact other geographical areas \cite{Adelman-1992-Energy,Liu-Chen-Wan-2013-EM}. Alterations in oil consumption, supply and demand, storage, and price fluctuations all contribute to shifts in the global trade pattern and the world economic situation \cite{Zhang-Ji-Fan-2015-EnergyEcon,Le-Chang-2013-EnergyEcon,Rafiq-Sgro-Apergis-2016-EnergyEcon}. Extreme events such as financial crises and the COVID-19 pandemic have led to economic recessions, triggering fluctuations in the energy market and trade shocks. The interdependence of trading systems makes it challenging for economies to remain isolated from these impacts. Consequently, these effects inevitably spread throughout the trade network to varying extents and may even disrupt the entire oil trade network, resulting the imbalance in global oil allocation. Therefore, exploring the resilience of the iOTNs in the face of extreme events is crucial. Proposing relevant measures to enhance network resilience holds significance for promoting stable development in the international oil trade while ensuring energy supply security for various economies.


Since Holling introduced the concept, scholars have extensively explored resilience \cite{Holling-1973-AnnuRevES}, spanning from engineering mechanics to ecosystem restoration research. Resilience can be defined as the ability of an entity to recover or rebound, signifying its capacity to return to its original state after external disturbances. Subsequent research has significantly broadened the connotations of resilience. The challenges confronting the global oil trade market are increasingly intricate and diverse, leading to a more nuanced understanding of the resilience of the oil trade network. The resilience can be viewed as the system's capability to maintain regular operation in response to various internal or external pressures, interferences, and disruptions while also possessing the ability to recover swiftly at the conclusion of such events. In essence, a highly resilient trade network necessitates both enduring shocks and swift recovery thereafter \cite{Vallina-LeQuere-2011-JTheorBiol,Allen-Angeler-Chaffin-Twidwell-Garmestani-2019-NatSustain}.


However, the assessment of the resilience of the iOTN is still in the initial exploration stage. Scholars mainly focus on studying the resistance of trade networks to extreme shocks based on historical data and simulation \cite{Du-Dong-Wang-Zhao-Zhang-Vilela-Stanley-2019-Energy,Ji-Zhang-Fan-2014-ECM,Gomez-Mejia-Ruddell-Rushforth-2021-Nature}, or measuring network resilience solely through static network metrics \cite{Chen-Chen-2023-JCleanProd,Shahnazi-Sajedianfard-Melatos-2023-EnergyRep}. Research on network recoverability still primarily exists in ecosystems \cite{Kirchner-Weil-2000-Nature,Hutchings-2000-Nature,Check-2005-Nature}. In recent years, some scholars have also introduced recoverability into transportation networks and proposed optimization strategies to enhance network recoverability \cite{Schwalm-Anderegg-Michalak-Fisher-Biondi-Koch-Litvak-Ogle-Shaw-Wolf-Huntzinger-Schaefer-Cook-Wei-Fang-Hayes-Huang-Jain-Tian-2017-Nature}. Nevertheless, there remains a need for additional development and enhancement in defining recoverability indicators and related research within energy trade networks \cite{Fair-Bauch-Anand-2017-SciRep,Foti-Pauls-Rockmore-2013-JEconDynControl}.


Previous research on the resilience of oil trade networks to extreme shocks, based on historical data or shock models, was insufficient for accurately and comprehensively measuring the network's resilience. This article commences by focusing on the iOTNs and introduces network efficiency indicators to assess the structure and function of the trade network. Subsequently, we develop an extreme event shock-recovery simulation model based on complex network method to simulate scenarios involving key economies in the trade network under extreme shocks of various intensities, as well as their recovery following the cessation of such shocks. Then, we measures resilience attributes of the iOTN from perspectives encompassing shock resistance and recoverability through defining relevant indicators. Finally, leveraging insights derived from the model and related results, attention is directed towards promoting sustainable development of the international oil trade. We propose future strategies for responding to extreme risks and providing decision-making references in order to mitigate energy trade vulnerability and enhance its recoverability.

The remainder of the paper is organized as follows. Section~\ref{S1:Data:Methodology} describes the data and methodology used in our analysis. Section~\ref{S1:Results} shows and explains the results of network resilience based on resistance and recoverability. Section~\ref{S1:Conclusion} includes the conclusion and discussion.

\section{Data and Methodology}
\label{S1:Data:Methodology}

\subsection{Oil trade data and construction of iOTNs}


The data is obtained from the United Nations Commodity Trade Database, managed by the United Nations Statistics Office, which provides global import and export data on commodities and services. The UN Comtrade database encompasses over 200 international and regional markets worldwide, covering more than 6,000 product categories with value data converted to US dollars based on the standard format of the United Nations Statistics Office. This database has been utilized in numerous scholarly research \cite{Yu-Jessie-Sharmistha-2015-Energy,Kharrazi-Fath-2016-EP,Du-Wang-Dong-Tian-Liu-Wang-Fang-2017-AEn,Zhong-An-Shen-Fang-Gao-Dong-2017-EP,Wei-Xie-Zhou-2022-Energy}. We use more complete data on oil import to analyze oil trade between economies \cite{Fan-Ren-Cai-Cui-2014-EconModel}.


The construction of the oil trade network relies on complex network methods, which can abstract the economies involved in trade data as network nodes, and the trade relationships between economies are network edges. We store the information of the iOTN in a $N\times N$ matrix $W=[w_{ij}]$, consisting of exporting economies, importing economies, and trading volumes, in order to facilitate the subsequent calculation experiment based on the extreme trade shock simulation model. Where $N$ is the number of economies in the iOTN, the matrix element $w_{ij}$ indicates that there is oil export trade between economies $i$ and $j$, and the trade volume is $w_{ij}$. Because of the significant lack of trade data in the last three years, the project studied the iOTNs from 1988 to 2022.

\begin{figure}[h!]
\centering
\includegraphics[width=0.99\linewidth]{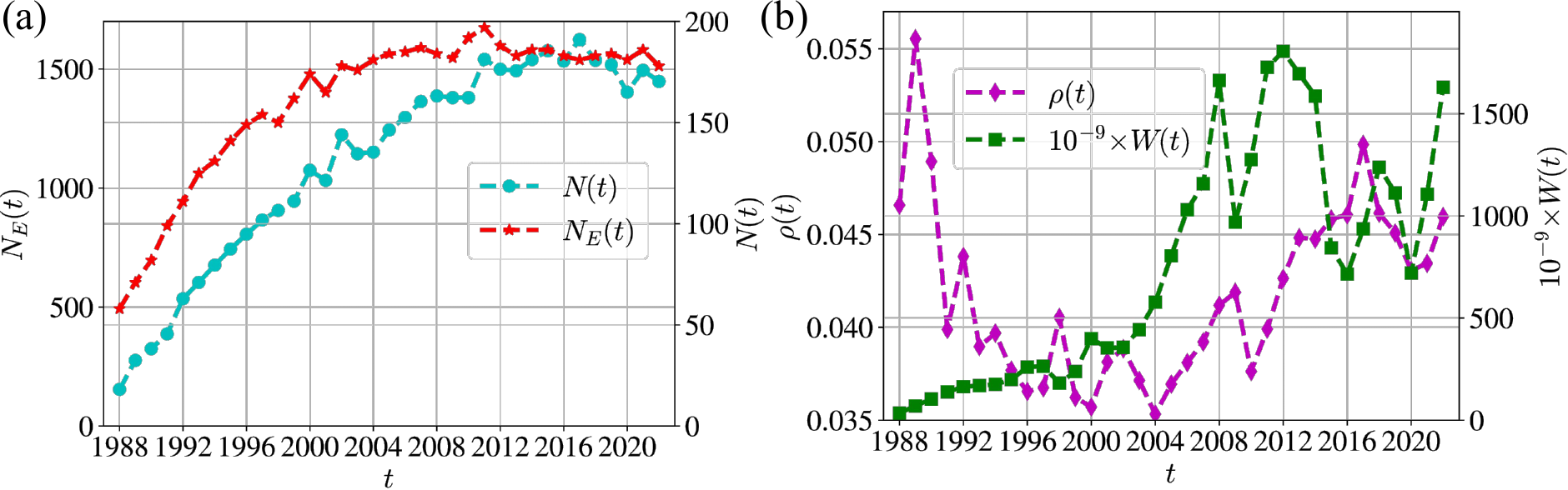}
  \caption{Evolution of the number of economies ($N$), relationships ($N_E$), trade volume ($W$), and network density ($\rho$) from 1988 to 2022.}
    \label{Fig:iOTN:NetStructure}
\end{figure}


Fig.~\ref{Fig:iOTN:NetStructure}(a) illustrates the evolution of the number of economies $N$ and trade relationships $N_E$ in the iOTNs from 1988 to 2022. The density of the trade network depicts the degree of trade connectivity between oil-trading economies and is calculated as equation $\rho = \frac{N_E}{N(N-1)}$. The evolution of the oil trade volume $W(t)$ and the network density $\rho$ are illustrated in Fig.~\ref{Fig:iOTN:NetStructure}(b). It can be seen that with the development of the global energy economy, more and more economies have joined the oil trading system, and the trade relationship between economies has become increasingly complex. Unlike the steady increase in economies and trade relations, the density of iOTNs and trade volumes are characterized by a certain degree of volatility. While both indicators show an overall upward trend, the volume of oil trade and network density show a significant downward trend after 1998, 2008, 2014, 2016 and 2020. These years correspond to major international events such as the Asian financial crisis, the global financial crisis, the U.S. shale oil revolution, and the sudden plunge in energy prices, which shows that trade shocks triggered by extreme events can spread to the entire trade network along with trade links, and the impacts on the structure of the iOTNs are significant. However, it is clear that, in most cases, extreme event shocks will not destroy the entire network, and the network structure and functionality will still be preserved to a certain extent, reflecting the fact that oil trade networks have certain risk-resistant properties. Besides, the values of attributes such as the trade volume and density of an economy gradually rise sometime after a sudden drop, suggesting that the iOTN has systemic recoverability, which reflects the fact that it is more reasonable to measure the resilience of oil trade networks in terms of shock resistance and recoverability dimensions.

\subsection{Network resilience assessment indicator}

\begin{figure}[h!]
\centering
\includegraphics[width=0.6\linewidth]{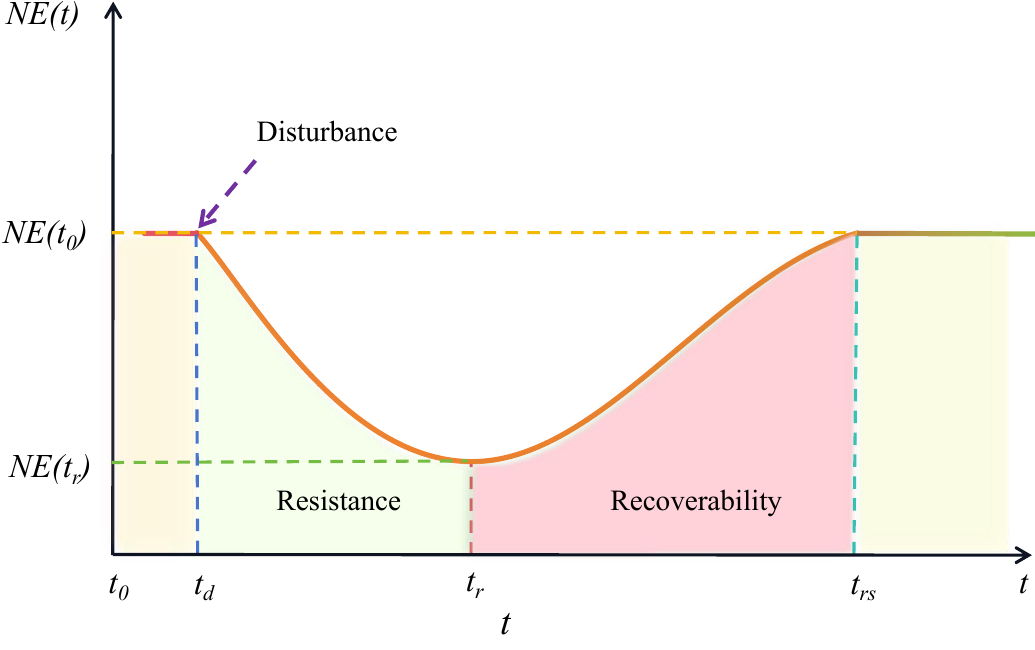}
  \caption{
Damage and recovery process of the iOTN under extreme shocks.
  }
    \label{Fig:iOTN:AttackAndRecovery}
\end{figure}

The resilience assessment in this study differs from previous measures of structural loss of networks under various types of shocks based solely on historical data. We refer to the ref \cite{Yu-Ma-Zhu-2023-ResourPolicy} to consider dynamic network resilience assessment, which divides the change process of network performance into two stages by simulating scenarios where external factors such as financial crises, regional armed conflicts, and other extreme events shock the network. The first stage involves a gradual reduction in network performance following the shock, during which we evaluate the resistance of the iOTNs. Subsequently, there is a recovery stage after the shock ends, ultimately returning to its initial state; based on this process, recoverability of the iOTN is assessed. Fig.~\ref{Fig:iOTN:AttackAndRecovery} illustrates schematic diagrams depicting changes in performance levels $NE(t)$ for impacted iOTNs at different stages of damage and recovery.

Both the traditional resilience assessment and the dynamic network resilience assessment in this paper require an examination of the change in network performance ($NE(t)$) as a basis for measuring resilience. Therefore, it is necessary to introduce a network performance assessment index. The strategic importance of oil resources and their specific distribution determine the adequacy of their global allocation, which holds significant implications for global energy governance and stabilizing energy supply and demand within individual economies. In assessing the performance of oil trading networks, we utilize network efficiency indicator to evaluate the efficiency of resource allocation \cite{Latora-Marchiori-2001-PhysRevLett,Xie-Wei-Zhou-2021-JStatMech}. This indicator $E^{W}$ can be applied to both unweighted or weighted networks, and in this study, we consider its introduction into the weighted iOTN $W_t$
\begin{equation}\label{Eq:Weight:Efficiency}
E^{W}=\frac1{N(N-1)}\sum_{i\neq j\in W_t}E_{ij},
\end{equation}
where $E_{ij}$is the path efficiency between economy $i$and economy $j$. In an unweighted network, $E_{ij}=\frac1{d_{ij}}$, the shorter the path length, the greater the efficiency. In the weighted network, the connected edges' weight should be considered. In a trade network, the weight is the value of oil trade between two economies. Therefore, the path efficiency between $i$and $j$ in the weighted network is
$E_{ij}=\frac1{\sum_{l\in L_{ij}}{\frac1{v_{l}}}}$,
Where $L_{ij}$is the edge set of the shortest path between the economies $i$ and $j$ in the weighted network $W_t$, $l$is the element in $L_{ij}$, and $w_{l}$ is the weight of the edge $l$. Therefore, the efficiency of the weighted network $W_t$ can be defined as
\begin{equation}\label{Eq:finial:weight:efficiency}
E=\frac1{N(N-1)}\sum_{i\neq j\in W_t}e_{ij}=\frac1{N(N-1)}\sum_{i\neq j\in W_t}\frac1{\sum_{l\in L_{ij}}{\displaystyle\frac1{w_l}}}.
\end{equation}
From Eq.~(\ref{Eq:finial:weight:efficiency}), network resource allocation efficiency increased with trade value $v$.

Due to economic development and changes in supply and demand, the volumes of oil entering the market may vary from year to year. To facilitate a clearer comparison of resource allocation efficiency across different years, the weights of the oil trade network can be normalized using $\bar{w}_l=\frac{w_l}{\langle w \rangle}$, where $\langle w \rangle$ represents the mean value of trade margin weight within the weighted oil trade network. This normalization process allows for standardizing formulas related to Eq.~(\ref{Eq:finial:weight:efficiency}) for assessing weighted oil trade network effectiveness:
\begin{equation}\label{Eq:weight:norm:efficiency}
E^{W}=\frac{E}{\langle w \rangle}.
\end{equation}
Then, the efficiency of resource allocation between different years is comparable.

Once the network performance indicator is determined, we can further interpret Fig.~\ref{Fig:iOTN:AttackAndRecovery}. It is evident that $t_{0}< t <t_{d}$ represents the initial stability stage of the oil trade network, where extreme events have not yet impacted the network. The stage from $t_{d}<t<t_{r}$ signifies when the network performance is compromised; it begins to be disrupted at $t_{d}$ and experiences a decline in performance, reaching its lowest value at $t_{r}$. Here, we introduce $R$ to denote the lowest point as well as the maximum impact it can withstand during this phase. Additionally, we utilize the change rate $ROC_{DS}$ to measure the speed of network performance decline during the impact stage
\begin{equation}\label{Eq:iOTN:R}
R=\min\{NE(t)\},\:t_{\mathrm{d}}<t<t_{\mathrm{rs}}
\end{equation}
\begin{equation}\label{Eq:iOTN:RDS}
ROC_{\mathrm{DS}}=\frac{NE(t_{i})-NE(t_{i}-\Delta t)}{\Delta t}
\end{equation}
where $NE(t)$ represents the discrete function of network efficiency varying with events; $r_{ts}$ indicates the time when the network returns to its initial stable state. $\Delta t$ represents the change in the event. We use the network resilience loss $LONE_{\mathrm{DS}} $to measure the resistance of the iOTN at the shock stage. $LONE_{\mathrm{DS}}$can be quantized as $NE(t)$ discrete curve and line $x=t_d$, $x=t_r$, Area enclosed by $y=NE(t_0)$:
\begin{equation}\label{Eq:iOTN:LONEDS}
LONE_{\mathrm{DS}}=\int_{t_{d}}^{t_{r}}(NE(t_{0})-NE(t))\mathrm{d}t,
\end{equation}
where $NE(t_0)$indicates the initial resilience level of the network, and $t_0$ indicates the initial stability time of the network. The $t_{r}< t <t_{rs}$ stage is the stage of the gradual recovery of network performance. Corresponding to the state of the impact stage, we introduce the change rate of network performance in the recovery stage $ROC_{RS}$, the loss of network performance in the recovery stage $LONE_{RS}$:
\begin{equation}\label{Eq:iOTN:ROC}
ROC_{_\mathrm{RS}}=\frac{NE(t_{i})-NE(t_{i}-\Delta t)}{\Delta t},t_{_\mathrm{r}}<t<t_{_\mathrm{rS}}
\end{equation}
\begin{equation}\label{Eq:iOTN:LONERS}
LONE_{\mathrm{RS}}=\int_{t_{r}}^{t_{rs}}(NE(t_{0})-NE(t))\mathrm{d}t.
\end{equation}

When $t \geq t_{rs}$, the network recovers, and the performance level of the network may be less than, equal to, or even higher than the initial level. This paper focuses on analyzing the situation in which the performance can recover to the initial level. To further evaluate the comprehensive resilience of the iOTN in the process of disturbance and recovery, the $Resilience^{W}$ comprehensive resilience value is introduced to measure:
\begin{equation}\label{Eq:iOTN:CRA}
Resilience^{W}=\int_{t_{d}}^{t_{r}}(NE(t_{0})-NE(t))\mathrm{d}t+\int_{t_{r}}^{t_{rs}}(NE(t_{0})-NE(t))\mathrm{d}t=LONE_{\mathrm{DS}}+LONE_{\mathrm{RS}}.
\end{equation}

\subsection{Extreme event shock and recovery simulation model in iOTN}

\begin{figure}[!t]
  \centering
  \includegraphics[width=0.7\linewidth]{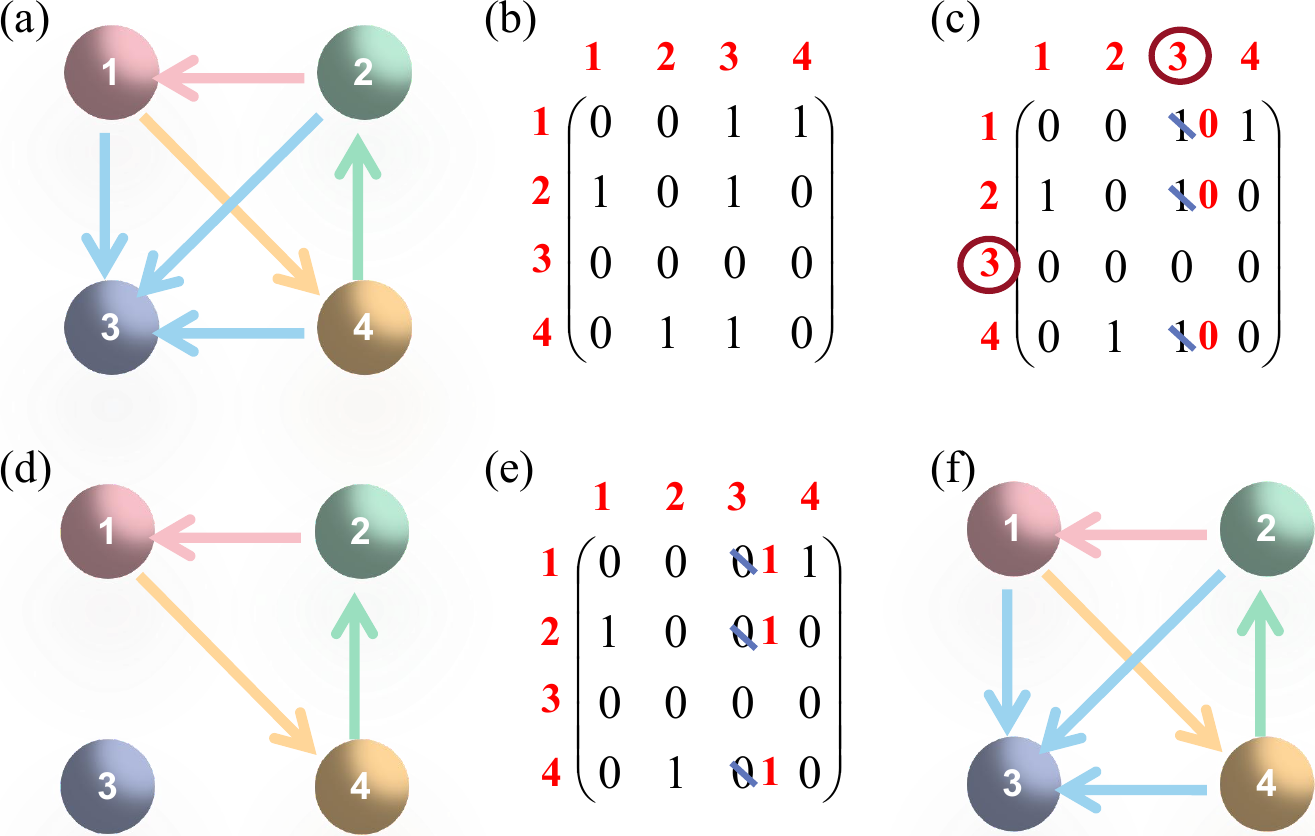}
  \caption{Schematic diagram of simulated attacks on economies.} 
  \label{Fig:iOTN:AttackModel}
\end{figure}

We cannot explore the trade network resilience property through discrete historical data. Therefore, it is necessary to construct a simulation model for extreme event shocks and recovery. Storing network information in a matrix allows us to model the negative impacts of extreme events on the iOTNs by deleting the values of specific matrix rows or columns that represent individual economies. Fig.~\ref{Fig:iOTN:AttackModel} shows an example of a numerical simulation of an extreme shock acting on a network. Fig.~\ref{Fig:iOTN:AttackModel}(a) is a simple example of a network that can be transformed into the matrix shown in Fig.~\ref{Fig:iOTN:AttackModel}(b). When the economy labeled "3" is shocked, the non-zero elements of the first row and column of the matrix can be assigned to 0, as shown in Fig.~\ref{Fig:iOTN:AttackModel}(c), and finally, the remaining network structure of economy "3" can be obtained as shown in (d), which is the remaining network structure of economy "3" after a complete shock. When the simulated trade system is recovered, we can restore the element of the trade relationship with the economy "1" to its original state, i.e., the element is given the original value in (e), and the deleted trade relationship is added back to the network. The final structure of the fully restored network is reflected in (f). Through numerical simulation models, we simulate various scenarios of shocks and recovery of the iOTN, explore the performance of the remaining network structure under each scenario, and ultimately assess the resilience attributes of the network through the constructed metrics.


The proposed numerical simulation model can simulate the situation when the shock occurs. The scale-free nature of the iOTN makes it harder to defend against targeted attacks. When extreme events shock critical economic nodes, it is highly likely to cause the entire network to fall apart \cite{Crucitti-Latora-Marchiori-2004-PhysRevE,Liu-Cao-Liu-Shi-Cheng-Liu-2020-Energy}. So, when applying the numerical simulation model, we simulate shocks to economies with significant trading positions or to trade relationships with high oil trade volumes.


First, the model is set to simulate an extreme shock to economies or trade relationships by removing 1\% of economies or trade relationships from the network, in descending order of their trade positions or volumes, over time $t$. Second, in the simulation, the performance of the iOTN reaches its lowest point when the top 50\% of economies are set to be unable to trade oil or when the top 50\% in terms of trade volume have a broken trade relationship. Once the performance is at its lowest point, we recover one by one, in the order of the economies and trade relations of the simulated shocks, until the network structure and performance are restored to their initial state. The trade position of an economy is measured using classical node centrality indicators, such as degree and betweenness, with detailed calculations in Ref \cite{Liu-Cao-Liu-Shi-Cheng-Liu-2020-Energy,Xie-Wei-Zhou-2021-JStatMech,Wei-Xie-Zhou-2022-Energy}.

\section{Empirical analysis}
\label{S1:Results}

\subsection{Evolving efficiency of iOTNs}
\begin{figure}[htp]
\centering
\includegraphics[width=0.5\linewidth]{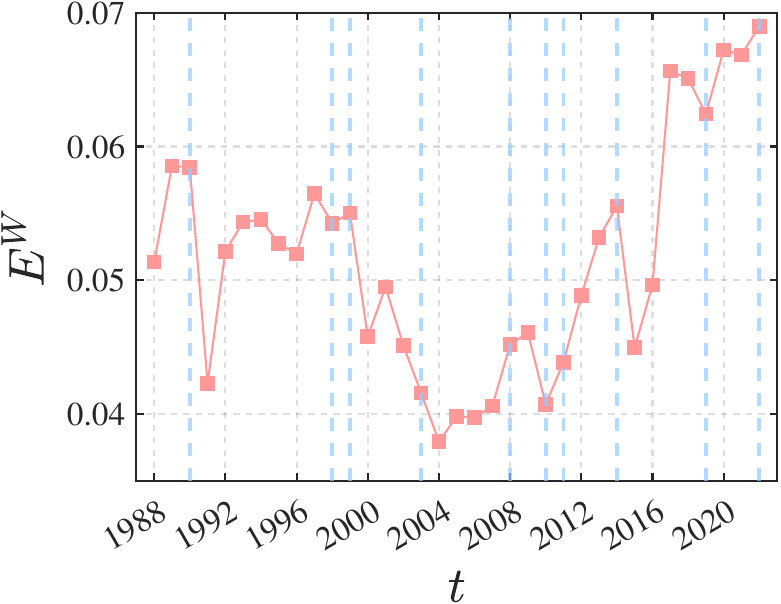}
\caption{Evolving efficiency of the iOTNs.}
\label{Fig:iOTN:Efficiency}
\end{figure}
Drawing from historical data, we conduct an analysis of the iOTN's resource allocation efficiency. The evolution of efficiency, taking into account trade flows, is depicted in Fig.~\ref{Fig:iOTN:Efficiency}. With the continuous adjustment of the global oil supply and demand structure, global economies have gradually established diversified trade relations, leading to a general upward trend in the allocation efficiency of oil resources.


There are notable declines following specific years such as 1997, 2008, 2014, 2016 and 2019. This indicates the significant impacts of extreme events such as the Asian financial crisis in 1997, the global financial crisis in 2008, the U.S. shale oil revolution in 2014, the sharp decline in oil prices in 2016, COVID-19 in 2019. Furthermore, the decrease in network efficiency after a major extreme event only persists until a certain critical point, beyond which network efficiency gradually rebounds.

\subsection{Trading status of oil trade economies}


The trade influences of economies in the iOTNs can be computed by the node influence measure indexes proposed in the methodology section. The findings indicate that an economy's trade influence fluctuates over time. For instance, we illustrate the top ten economies with the most significant trade influence under different measurement indicators of the iOTN for 2020 in Table~\ref{tab:2020:node:criticalityRank10}.

\begin{table*}[htp]
  \centering
  \caption{Top 10 economies identified based on different influence indicators in 2020}
    \scalebox{0.66}{
    \begin{tabular}{ccccccc}
    \toprule
    Rank & PageRank & Outdegree & Indegree & Outcloseness & Incloseness & {Betweenness} \\
    \midrule
    1     & Netherlands & USA   & Netherlands & USA   & Netherlands & Netherlands \\
    2     & Singapore & Russian Federation & ASEAN & Russian Federation & ASEAN & USA \\
    3     & ASEAN & Nigeria & Singapore & Nigeria & Singapore & United Kingdom \\
    4     & Spain & Saudi Arabia & China & Saudi Arabia & France & United Arab Emirates \\
    5     & France & United Kingdom & India & Norway & India & France \\
    6     & South Africa & Norway & Spain & United Kingdom & Spain & South Africa \\
    7     & Rep. of Korea & Azerbaijan & France & Iraq  & Rep. of Korea & Spain \\
    8     & India & Iraq  & Rep. of Korea & Algeria & Thailand & Kazakhstan \\
    9     & China & Kazakhstan & Italy & Libya & China & Rep. of Korea \\
    10    & United Arab Emirates & United Arab Emirates & USA   & Kazakhstan & Germany & Belgium \\
    \midrule
    Rank & Authorities & Hubs & Clustering & Within-module   & Outside-module   & Participation \\
    \midrule
    1     & ASEAN & USA   & Myanmar & USA   & USA   & USA \\
    2     & Netherlands & Russian Federation & Paraguay & South Africa & Netherlands & Canada \\
    3     & Singapore & Nigeria & Togo  & United Kingdom & China & France \\
    4     & China & Saudi Arabia & Luxembourg & Netherlands & South Africa & Brazil \\
    5     & India & Norway & Trinidad and Tobago & Rep. of Korea & ASEAN & Cote d'Ivoire \\
    6     & France & Algeria & Chad  & Singapore & United Kingdom & Portugal \\
    7     & Spain & Iraq  & Guatemala & Russian Federation & Nigeria & Spain \\
    8     & Rep. of Korea & United Kingdom & Guyana & Spain & Russian Federation & Norway \\
    9     & Italy & Azerbaijan & Sri Lanka & United Arab Emirates & India & South Africa \\
    10    & USA   & Libya & Finland & India & Singapore & Nigeria \\
    \bottomrule
    \end{tabular}}%
  \label{tab:2020:node:criticalityRank10}%
\end{table*}%

In Table.~\ref{tab:2020:node:criticalityRank10}, the influence of economies such as the United States and Netherlands is more substantial both at the local or global network structure and at the level of the association structure. In addition, emerging economies, represented by China, have shown their importance in the iOTNs. There is some similarity in the 2020 measurements. However, most indicators are not measured the same way, suggesting that there may be some correlation between the different indicators that need further exploration. Influence indicators based on local structure can screen out major energy-producing and consuming economies that are more focused on considering the volume rather than the distribution of import-export trade relationships, as Netherlands and the United States. The clustering indicator filters out economies with closer energy cooperation on a small scale, which may be small and have only a localized influence. Besides, the three influence measures based on the modular structure consider the regional agglomeration characteristics of oil trade, capturing the cross-regional and cross-module trade behavior of an economy to identify the expansion of an economy's sphere of influence. The results of applying Within-module and outside-module indicators to measure economic influence are more similar.

\subsection{Simulation of shocks to individual economy and trade relationship}

\begin{figure}[!t]
\centering
\includegraphics[width=0.95\linewidth]{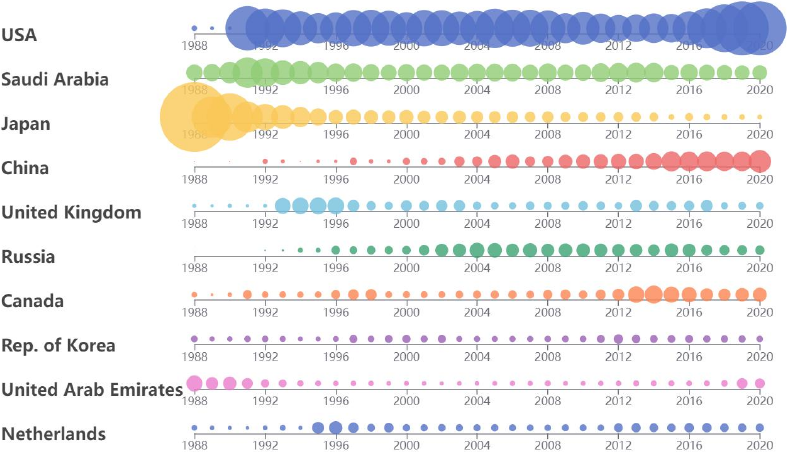}
\caption{
The top 10 economies with the most significant declines in network efficiency under the individual economy shock simulations.
}
\label{Fig:iOTN:Nodes:Critop10:2020}
\end{figure}
\begin{figure}[!t]
\centering
\includegraphics[width=0.95\linewidth]{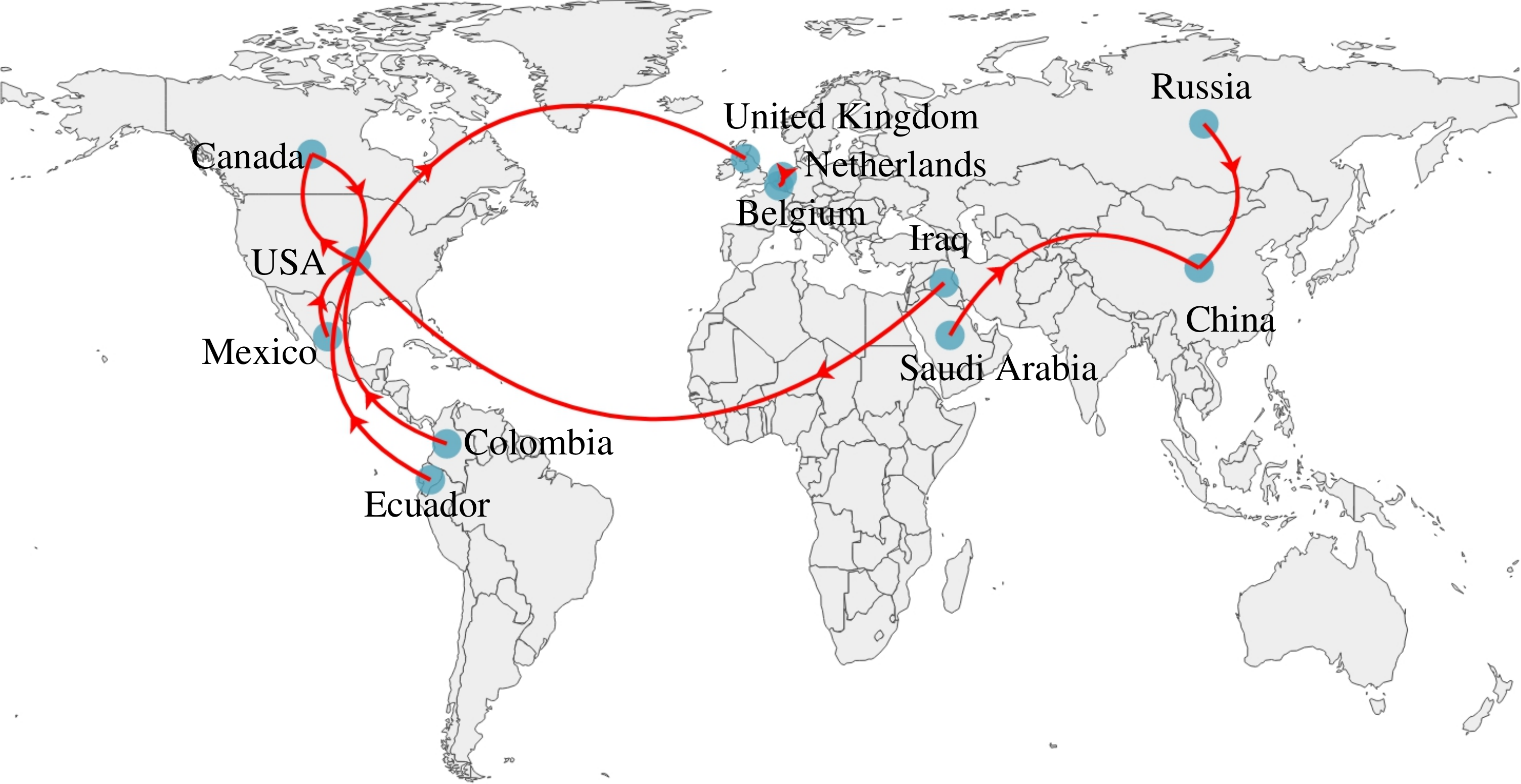}
\caption{
The top 10 trade relationships with the most significant decrease in network efficiency under single trade relationship shock simulation.
}
\label{Fig:iOTN:Edges:Critop10:2020}
\end{figure}

Economies and trade relationships in the iOTNs are heterogeneous, occupying different positions. The extent of decline in efficiency will also vary if specific economies or trading relationships are removed from the network. Based on the network efficiency, we analyze the degradation of network performance and present final results based on the magnitude of decrease in network efficiency when individual economies and trade relations experience shocks. Fig.~\ref{Fig:iOTN:Nodes:Critop10:2020} and Fig.~\ref{Fig:iOTN:Edges:Critop10:2020} illustrate the simulation results.

Fig.~\ref{Fig:iOTN:Nodes:Critop10:2020} depicts the top ten economies the most significant overall network performance degradation after shocks to economies over the years. In the figure, economies such as the United States, Saudi Arabia, and Japan play pivotal roles in resource allocation across all years. Post-1990, the United States supplanted Japan as the most critical economy in the iOTN. The United States has consistently been identified as the most crucial economy in numerous years, aligning with its status as a leading developed economy. In recent years, global oil dynamics have become increasingly intricate, with the United States transitioning from being a major oil importer to achieving energy self-sufficiency. This shift is attributed to complex geopolitical relations and ongoing regional conflicts that have led to a substantial reduction in U.S. dependence on traditional oil-producing regions in the Middle East and Africa \cite{An-Wang-Qu-Zhang-2018-Energy}. Nevertheless, its dominant position in energy trade markets remains undeniable. Furthermore, large energy-consuming countries like China also hold essential positions in energy resource allocation.


Due to the extensive number of trading relationships in the iOTN, the disparities in simulated shock results are minimal. To provide a more visual representation of these trade relationships, we map out and display the top ten key trade relationships in 2020 as depicted in Fig.~\ref{Fig:iOTN:Edges:Critop10:2020}. Notably, large energy-resource countries such as Iraq and Saudi Arabia are effectively identified when considering both network structure and trade weights in the network efficiency indicator. The trade relations involving economies like the U.S., Canada, and Netherlands are consistently prominent across various results, aligning with previous economy shock simulations. Consequently, it is evident that global energy resource allocation and performance in the iOTN are closely intertwined with significant energy-producing and consuming countries along with their respective trade relations. In efforts to mitigate and respond to energy risks, attention should be directed towards these economies and their trade relations.

\subsection{Shock-recovery simulation results of iOTN}


\begin{figure}[!ht]
\centering
\includegraphics[width=0.95\linewidth]{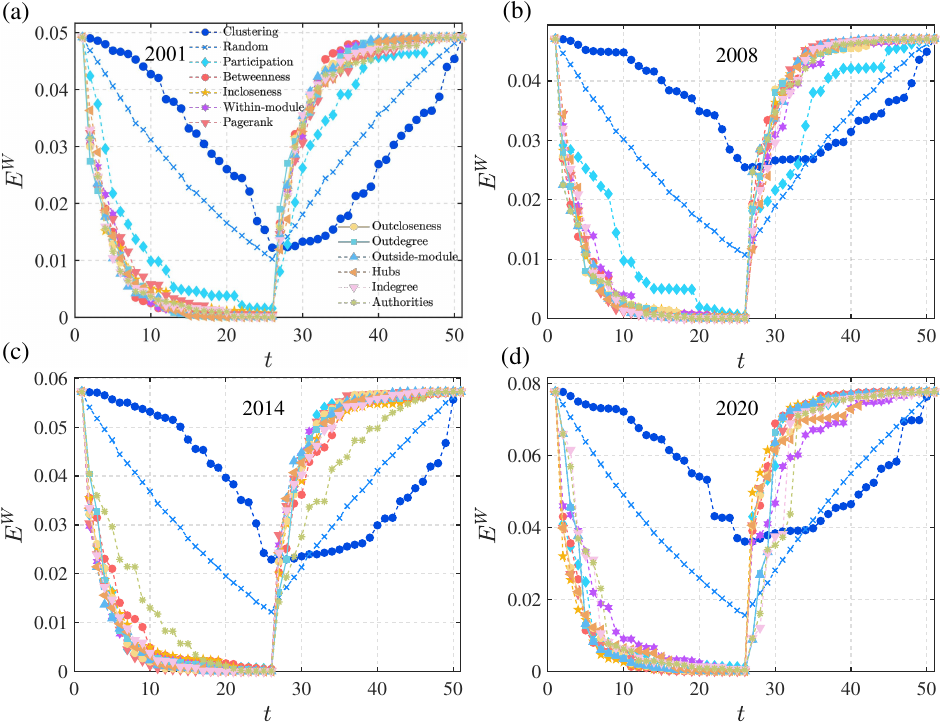}
\caption{
Changes in the efficiency of iOTNs under economic shocks and recoveries.
}
\label{Fig:iOTN:Economy:Attack:WeightedEfficiency}
\end{figure}

First, Fig.~\ref{Fig:iOTN:Economy:Attack:WeightedEfficiency} illustrates the change in network performance after an extreme event shock in economies with crucial trading positions. We have selected four years' simulation results in 2001, 2008, 2014, and 2020 for presentation; similar results were obtained for other years. The figure demonstrates that as time $t$ progresses, the broader the shock scope of extreme event on economies holding significant global trading positions, the more inefficient resource allocation becomes in the iOTNs. Under simulated shock strategies based on various influence indicators apart from local clustering coefficients, all decreasing trends in network performance exhibit cliff-like patterns. This contrasts with the linear decreasing trend observed in our random shock simulation control group and further underscores the significance of key economies in global oil resource allocation. The local clustering coefficients yield different outcomes across strategies, indicating a robust localization effect when portraying economies' position. Additionally, disruption to very small trading blocs has minimal impact on overall network resource allocation efficiency.


When the network performance reduce to the lowest point, we continue to simulate trade recovery after the shock based on the influence measures of economies. The results indicate that once the trade relations of top-ranked trading economies are restored after the shock, there will be a significant improvement in the performance of the iOTNs. Conversely, any enhancement in allocation efficiency for oil resources among lower-ranked trading economies is expected to be minimal.


The simulation results of economies exposed to extreme shocks align with the scale-free nature of the iOTNs, where a small number of economies have a large number of trading relationships \cite{Liu-Cao-Liu-Shi-Cheng-Liu-2020-Energy}. The structure and operation of the iOTNs are significantly disrupted when economies with substantial trade influence and numerous trading relationships experience shocks. Consequently, the swift restoration of trade relations among such economies after a shock is crucial for overall recovery in the iOTN. The differences in the trade status measurement indicators of various economies in the simulation results, we make further comparative explanations in the resilience analysis.

\begin{figure}[!ht]
\centering
\includegraphics[width=0.95\linewidth]{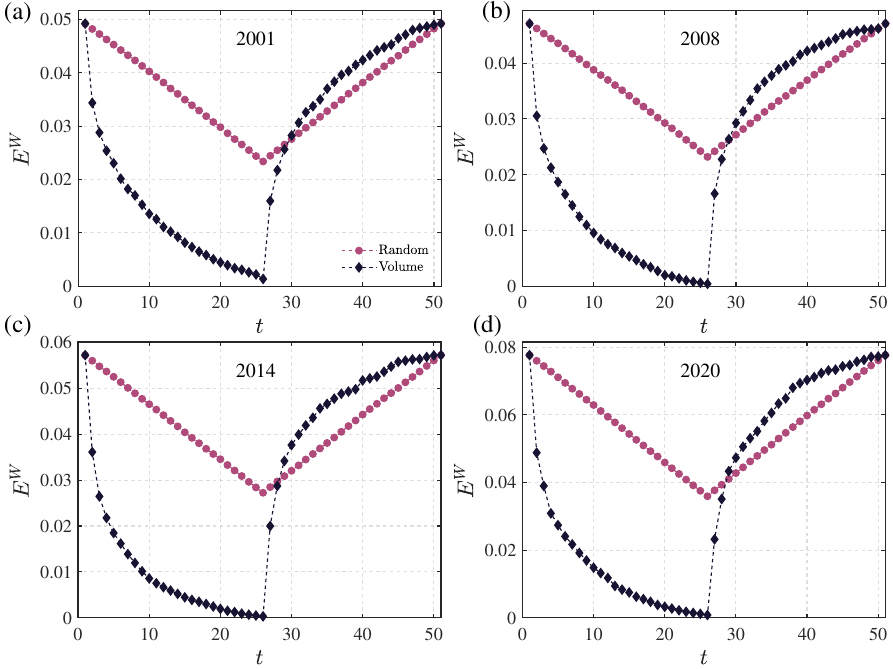}
\caption{
Changes in the efficiency of iOTNs under shocks and recoveries of trade relationships.
}
\label{Fig:iOTN:Relationship:Attack:WeightedEfficiency}
\end{figure}



Furthermore, as depicted in Fig.~\ref{Fig:iOTN:Relationship:Attack:WeightedEfficiency}, we present a scenario where an extreme event impacts critical trade relationships. As the shock widens in scope, disruption occurs, then the severe impact diminishes and the trade relationship resumes over time. Various previous indicators have been used to describe the importance or centrality of trade relations. In the simulation framework, we initially introduce the most intuitive trade volume to assess the significance and status of trade relationships. The figure illustrates that trade relationships with high volumes are among the first to suffer damage from shocks, resulting in significant degradation of network performance compared to linear variation results obtained from stochastic shock-recovery simulations. When comparing these results with simulations on recovery from shocks affecting economy nodes, it is evident that network performance degradation and recovery curves after trade relationship shocks are much flatter due to all of an economy's trade relationships being impacted and disrupted by a shock. In contrast, disruptions in individual trade relationships are more targeted. For equivalent levels of exposure to shocks, an economy's exposure leads to more substantial loss within the network in terms of functionality and structure.

\subsection{Resilience of iOTNs}

\begin{figure}[!t]
\centering
\includegraphics[width=0.9\linewidth]{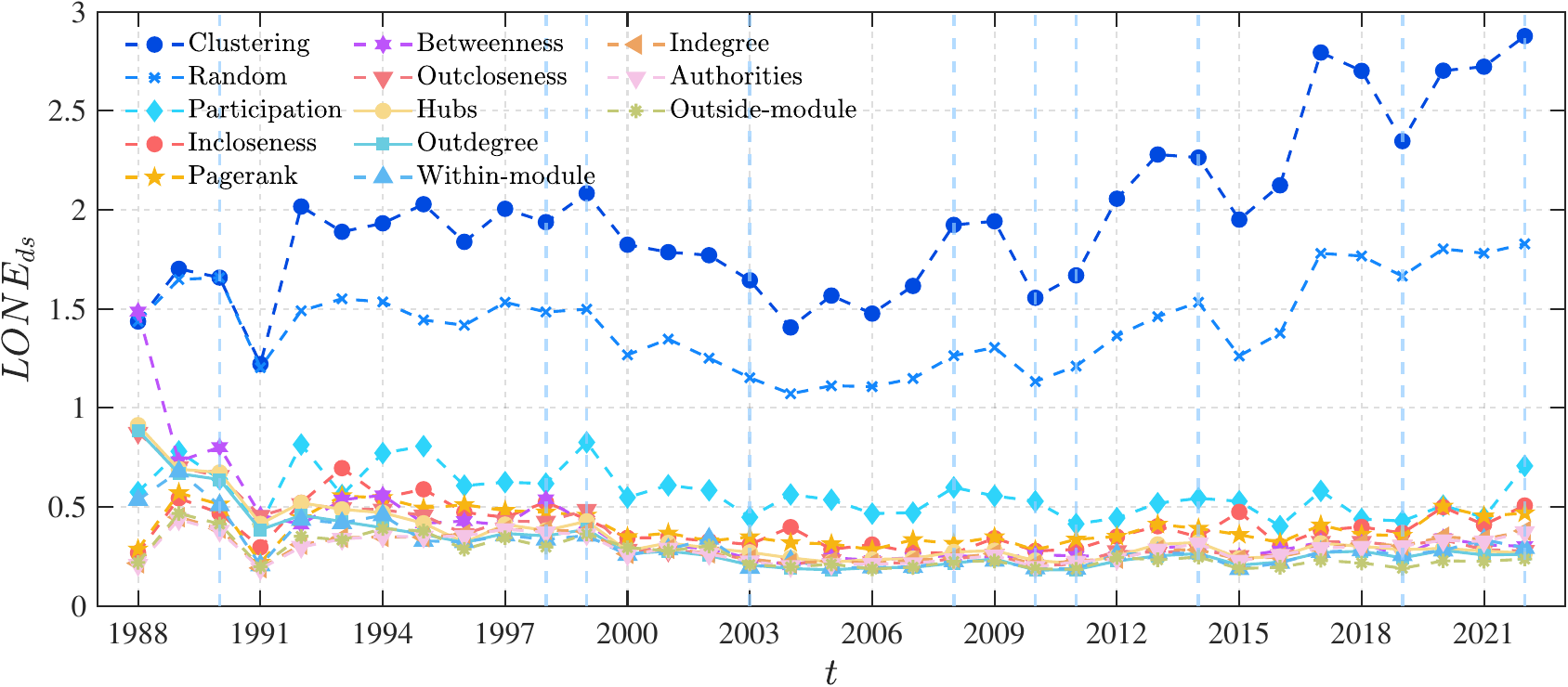}
\caption{
Resistance evolution of iOTNs under economic shocks. The light blue vertical dashed lines mark the years of major international events.
}
\label{Fig:Nodes:Resistance:W:Evolution}
\end{figure}
\begin{figure}[!t]
\centering
\includegraphics[width=0.9\linewidth]{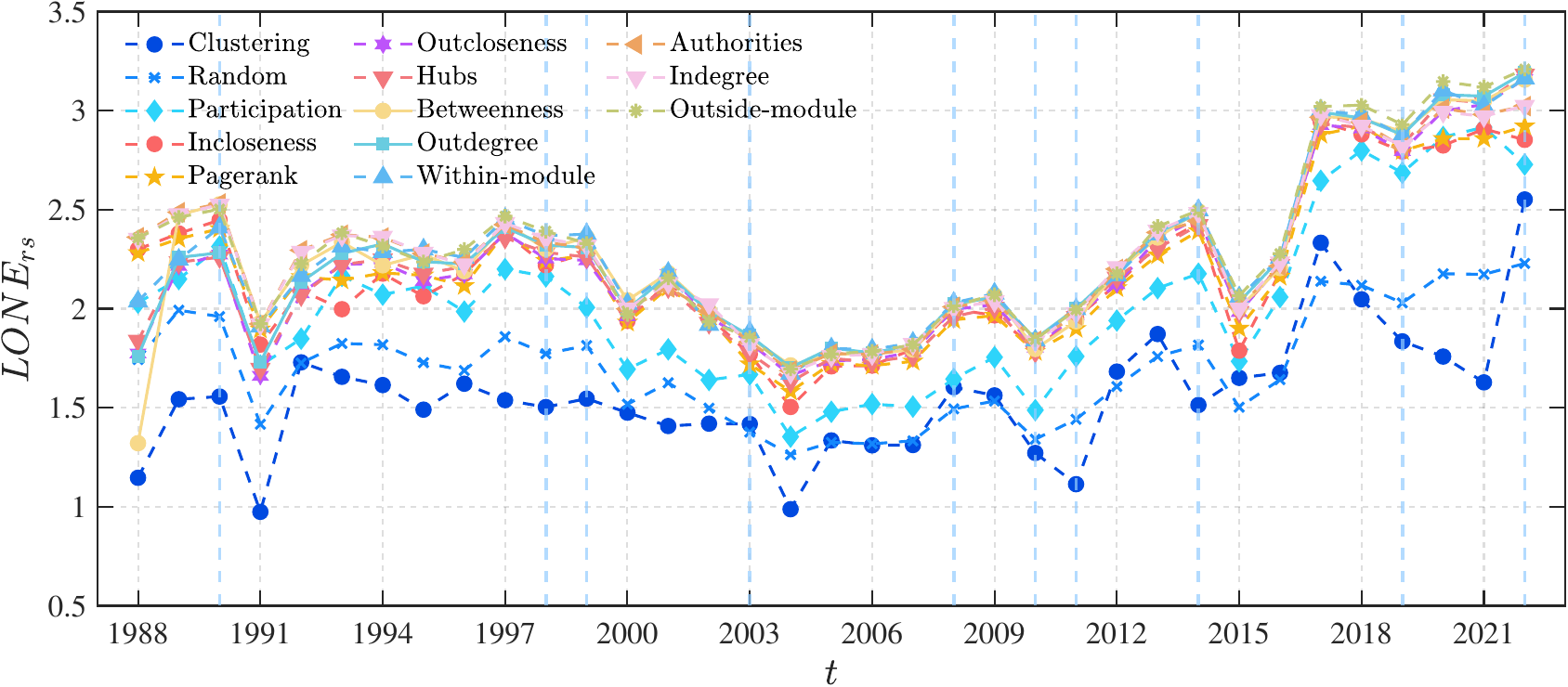}
\caption{
Recoverability evolution of iOTNs under economic shocks.}
\label{Fig:Nodes:Recoverability:W:Evolution}
\end{figure}

Utilizing the resilience indicators outlined in the methodology section and the results of simulated shocks, we can assess the shock resistance and recoverability of the iOTN under various simulation scenarios involving shocks to economies and trade relations, which allows for a comprehensive evaluation of resilience. Initially, we examine the evolution of shock resistance and recoverability in the iOTNs under simulations of economic shocks, as depicted in Fig.~\ref{Fig:Nodes:Resistance:W:Evolution} and Fig.~\ref{Fig:Nodes:Recoverability:W:Evolution}. The years corresponding to historically significant international events are denoted by blue vertical lines in these figures. The trend indicates a substantial decrease in shock resistance and recoverability following extreme events, signifying significant impacts on the performance of the iOTNs.

The results of resistance evolution reveal that the resistance of the iOTN remains relatively stable. Simulation outcomes based on the trade status of economies exhibit minor variations from year to year, with only the resistance derived from local clustering coefficients showing a higher level, surpassing even the results obtained from stochastic shocks. This suggests that economies identified as having high trade status by local clustering coefficients possess strong local cooperative influence, and their relationship ruptures have minimal impact on global oil allocation. In contrast, simulation results for other trade status measures are notably lower than those from stochastic simulations in terms of resistance. This indicates that vulnerability in the iOTN becomes more pronounced when extreme shocks affect economies with numerous trade relations and substantial trade influence. It also underscores that trade agglomeration continues to be a significant source of vulnerability for the iOTNs.

Further examination of the recoverability evolution under economic impacts is presented in Fig.~\ref{Fig:Nodes:Recoverability:W:Evolution}, where there is a more distinct discernibility in the recoverability evolution compared to the stability and similarity observed in the impact resistance results. Notably, simulation outcomes for recovery based on modular externality demonstrate the highest level, indicating that prioritizing the recovery of economies with higher modular externality and their trade relations leads to better recovery within the iOTN. Conversely, prioritizing the recovery of economies with high values of local clustering coefficients proves to be less effective in restoring efficiency in global oil resource allocation. Additionally, an intriguing conclusion can be drawn from all simulation scenarios - there is an overall increasing trend in recoverability in the iOTNs, which suggests that the network may be evolving into a more optimized structure capable of recovering more rapidly from unexpected shocks.

\begin{figure}[!t]
\centering
\includegraphics[width=0.9\linewidth]{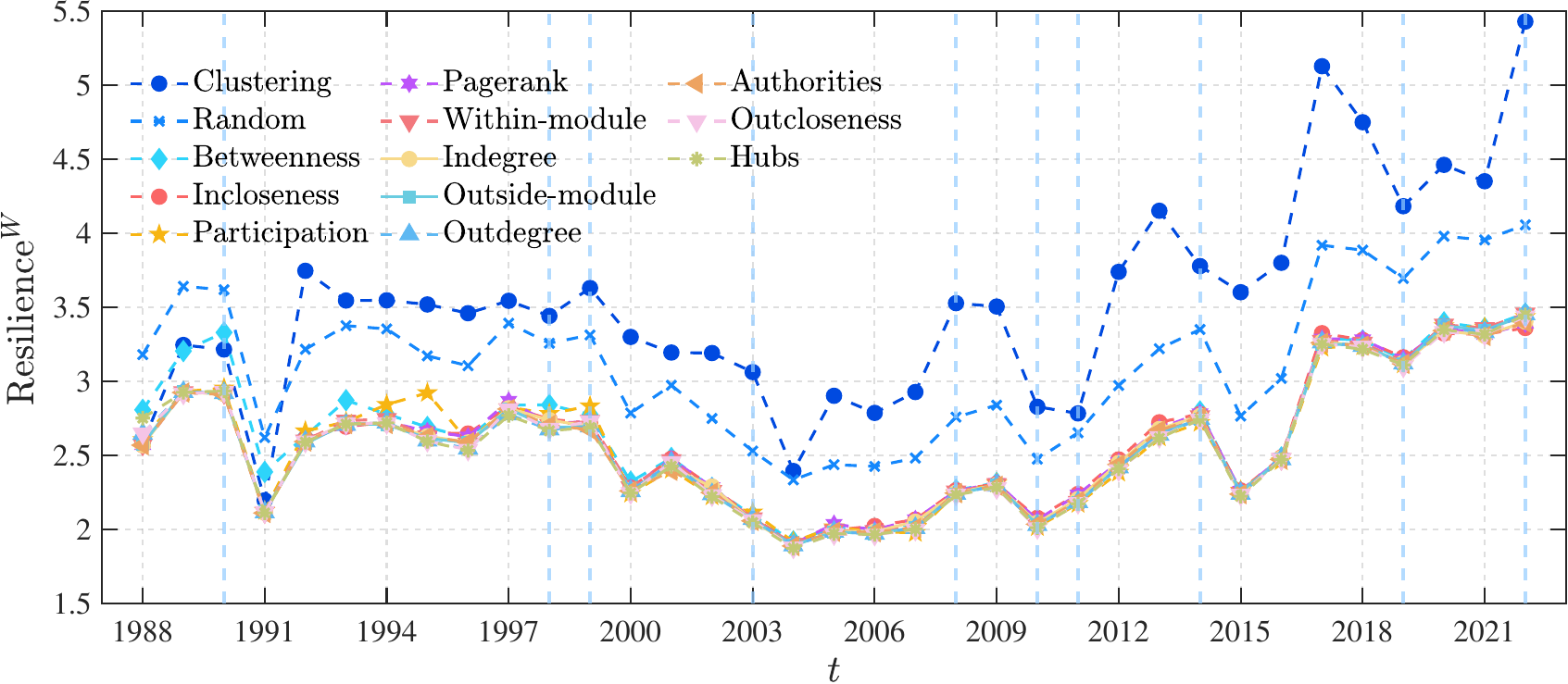}
\caption{
Resilience evolution of iOTNs under economic shocks. }
\label{Fig:Nodes:Resilence:Evolution:W}
\end{figure}


Following the analysis of shock resistance and recoverability within the iOTN, we can comprehensively evaluate its resilience according to Eq.(\ref{Eq:iOTN:CRA}). Subsequently, we present the resilience evolution based on simulations of the shock-recovery model involving economies in Fig.~\ref{Fig:Nodes:Resilence:Evolution:W}.



\begin{figure}[htp]
\centering
\includegraphics[width=0.9\linewidth]{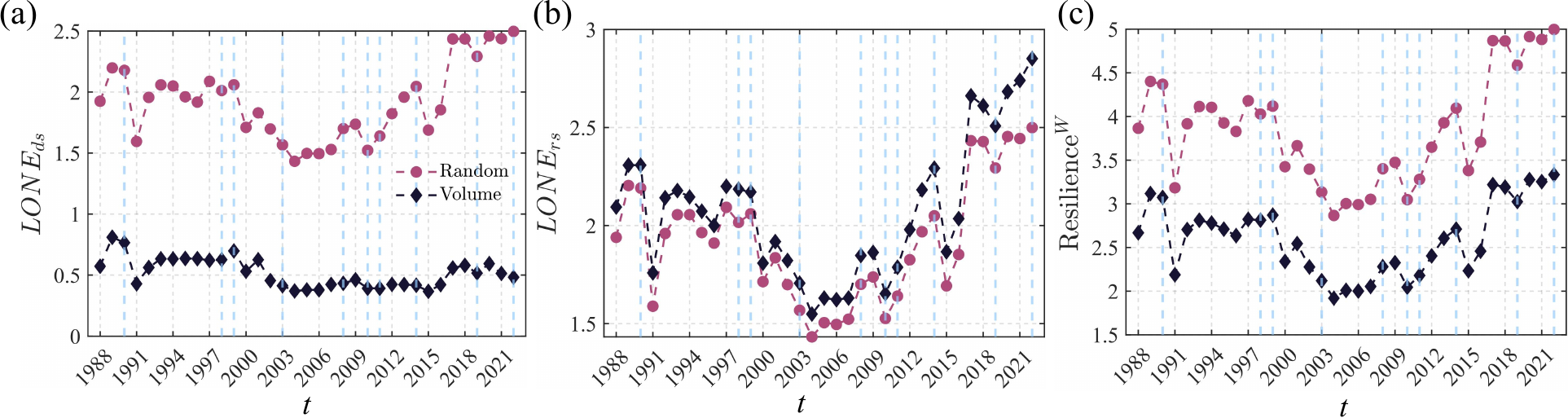}
\caption{
Resistance and recoverability evolution of the iOTN under shocks of trade relationships.
}
\label{Fig:Edges:Resistance:Evolution}
\end{figure}

Based on the shock-recovery simulation of trade relations involving economies with oil trade volumes, we observe the evolution of shock resistance and recoverability within iOTNs under trade relationship shocks in Fig.~\ref{Fig:Edges:Resistance:Evolution}(a) and (b). The shock resistance depicted in Fig.~\ref{Fig:Edges:Resistance:Evolution}(a) is notably lower than the results from stochastic shock simulations under the trade volume-based shock scenario, indicating that trade relationships with larger volumes play a crucial role in global oil resource allocation. In Fig.~\ref{Fig:Edges:Resistance:Evolution}(b), the recoverability results are less consistent with our expectations, and there is only a minor difference between the evolution results under both simulation scenarios. This can be attributed to the scale-free nature of iOTNs, where only a small number of trade relationships involve significant oil flows while most carry smaller flows. Consequently, when a few relationships with smaller flows are restored, their impact on resource allocation efficiency is relatively limited. The response of iOTNs to extreme event shocks remains evident in the results of trade relationship shock-recovery simulations as shown in Fig.~\ref{Fig:Edges:Resistance:Evolution}(c).


In Fig.~\ref{Fig:Nodes:Resilence:Evolution:W} and Fig.~\ref{Fig:Edges:Resistance:Evolution}(c), we observe an upward trend in the resilience of the iOTN based on shock-recovery simulations involving both economies and trade relationships. The evolution of resilience under various scenarios, derived from shock-recovery simulations of economies' trade positions, exhibits similar numerical values and evolutionary trends.

Apart from the distinct results related to local clustering coefficients, the resilience observed in other shock-recovery scenarios is notably lower than that of random simulations. Shock-recovery simulation results based on trade volumes significantly differ from those of the random control group, with trade relationships carrying higher volume displaying lower resilience values.

Economies and trade relationships holding key trading positions are crucial for global resource allocation. The evolution of resilience indicates a substantial decline in the index during corresponding or adjacent years to extreme events, signifying significant impacts on the performance of the iOTNs. The full manifestation of this impact may exhibit some delay. The increasing trend in resilience indicators reflects the ability of the oil trade network to optimize its structure and function.

\section{Discussion and application}
\label{S1:Conclusion}


"Resilience" denotes the capacity of a system to evade state transitions or swiftly recover from exceptional states, enabling the system to adapt its operations to uphold essential functions in the face of failures or disturbances. With advancements in network science theories, scholars have increasingly delved into exploring complex network systems and their resilience in real-world \cite{Liu-Li-Ma-Szymanski-Stanley-Gao-2022-PhysRep}. Against this backdrop, our research focuses on examining the resilience of the iOTNs. A highly resilient iOTN can help economies engaged in oil trade minimize the shock of extreme contingencies and expedite recovery from such events. High resilience encompasses both high resistance and recoverability, constituting an integral component of energy security assessments in international energy markets.


In this paper, we initially construct the iOTNs. Subsequently, to quantitatively assess the resilience of the iOTNs, we overcome the limitations of discrete historical data by developing a simulation model for shock-recovery simulations within the iOTNs. Simulating scenarios representing various types of extreme shocks experienced by the iOTNs based on the model. Then, we investigate alterations in both structure and function in each scenario and establish indices for network shock resistance and recoverability. Ultimately, we construct a resilience assessment index for the international oil trade system based on resistance and recoverability indicators to evaluate resilience attribute of the iOTNs.


The international oil trading system has grown increasingly intricate with the evolution and transformation of energy trade. As more economies participate in the globalization process, they become intricately interconnected through complex trade relationships. Key attributes of the iOTN, such as network density and trade volume, exhibit an overall upward trajectory. However, events like financial crises, energy revolutions, extreme fluctuations in the energy market, and other major international emergencies have significantly impacted the structure of the iOTN. These events have caused a certain degree of damage to its structure and function; however, these effects are gradually being restored over time. This restoration reflects the resilience attribute of the international oil trade system and validates measuring its resilience based on dimensions such as shock resistance and recoverability.


Prior to constructing the shock-recovery simulation model, we introduce network efficiency as a performance measure of the iOTNs. The empirical analysis of network efficiency evolution demonstrates its ability to effectively depict the structure and function of the network from a resource allocation perspective. Global oil resource allocation efficiency has shown improvement over time; however, major international events and extreme occurrences have had significant impacts on efficiency. The results from simulated shocks and recovery align with the scale-free properties of the trade network. When subjected to extreme contingencies, economies with numerous trade relationships, pivotal trade positions, or high-volume trade relations experience a substantial reduction in network efficiency. Furthermore, a certain degree of trade agglomeration contributes to shock transmission along these trade relationships, serving as an essential source of vulnerability for the network. Simulation results for recovering from network shocks underscore that restoring economies with crucial positions and trade relations is imperative for enhancing network performance.


Based on the analysis of resistance, recoverability, and resilience indicators presented in this paper, it is evident that both the resistance and recoverability of the network exhibit a declining trend during years marked by extreme events. The resistance of the network under simulation based on influence indicators of economies does not display a significant trend over increasing years. Notably, there is a discernible upward trend in the evolution of recoverability in the iOTNs, indicating its gradual development into a system with an increasingly optimized structure. In comparison to the early stages of energy market development, it is apparent that the international oil trade system is maturing gradually. This maturity enables it to more effectively manage unforeseen event impacts.


Based on these findings, we propose the following management recommendations from the perspective of global energy governance and energy security maintenance: Firstly, there should be an emphasis on resilience policies in the global energy governance system to reduce vulnerability to single events by diversifying trade corridors and establishing flexible supply chains. Secondly, energy economies should take proactive measures to mitigate potential shocks by reducing their reliance on specific regions or supplying economies, particularly those with significant trading positions. This can be achieved through vigilance regarding extreme contingency risks, real-time awareness of market changes, and the development of detailed crisis response plans to ensure swift implementation of appropriate measures in case of contingencies. Furthermore, investments in emerging technologies and renewable energy sources are essential for enhancing the sustainability of energy systems. Lastly, participants in the energy market should prioritize stable trade relationships with substantial resource allocations while establishing reliable supply chains and market forecasts to minimize market uncertainty.


Energy security is crucial not only due to its direct impact on national economies and security but also for ensuring stability. A dependable oil supply is fundamental for sustaining normal operations within transportation, industry, agriculture as well as promoting a sustainable transition towards clean energy and environmental protection. Implementing these measures will contribute towards enhancing overall resilience in the iOTNs while upholding energy security standards that enable robust operation amidst challenges.


Network topology is one of the important factors that determine the resilience of the iOTNs, and this study offers a comprehensive assessment of the network's structural resilience to provide insights for optimizing oil resource allocation and enhancing trade network resilience. However, the network performance indicators for measuring structural resilience still have potential for further optimization. Future research will delve into exploring additional network topological characteristics, political factors, transaction costs, and other conditions that influence the evaluation of network performance.

\section*{Acknowledgment}

This work was supported by Scientific and Technological Research Program of Chongqing Municipal Education Commission, and the Fundamental Research Funds for the Central Universities.

\section*{Data Availability}

Oil data sets related to this article can be found at https://comtrade.un.org/, an open-source online data repository. 


\end{document}